\documentclass[twocolumn,showpacs,preprintnumbers,amsmath,amssymb,floatfix]{revtex4}
\usepackage{graphicx}

\begin{document}

\title{Nanogaps with very large aspect ratios for electrical measurements}

\author{A. Fursina$^{1}$, S. Lee$^{2}$, R.G.S. Sofin$^{3}$, I.V. Shvets$^{3}$, D. Natelson$^{2, 4}$}

\affiliation{$^{1}$ Department of Chemistry, Rice University, 6100 Main St., Houston, TX 77005}
\affiliation{$^{2}$ Department of Physics and Astronomy, Rice University, 6100 Main St., Houston, TX 77005}
\affiliation{$^{3}$ CRANN, School of Physics, Trinity College, Dublin 2, Ireland}
\affiliation{$^{4}$ Department of Electrical and Computer Engineering, Rice University, 6100 Main St,.Houston, TX 77005}

\date{\today}


\begin{abstract} 
For nanoscale electrical characterization and device fabrication it is often desirable to fabricate planar metal electrodes separated by large aspect ratio gaps with interelectrode distances well below 100 nm. We demonstrate a self-aligned process to accomplish this goal using a thin Cr film as a sacrificial etch layer. The resulting gaps can be as small as 10 nm and have aspect ratios exceeding 1000, with excellent interelectrode isolation. Such Ti/Au electrodes are demonstrated on Si substrates and are used to examine a voltage-driven transition in magnetite nanostructures. This shows the utility of this fabrication approach even with relatively reactive substrates.
\end{abstract}

\maketitle

There is much interest in the electronic characterization of nanoscale
materials and the creation of working molecular-based devices
\cite{MolEl_review_2000}.  Both goals demand the fabrication of
metallic electrodes separated by a distance comparable with the
targeted length, {\it i.e.} a few nanometers.  Much recent progress
has been made in nanogap fabrication, and several techniques
were proposed, including electromigration
\cite{EM1_1999,EM2_2005,EM3_2002}, electrodeposition
\cite{ED1_2002,ED2_2000}, mechanically controlled break junctions
\cite{BreakJunc_Tour_1997}, advanced e-beam lithography methods
\cite{Ebeam_align_2000,Ebeam_TEM_2007,Ebeam_OverExp_2002,Ebeam_UnderExp_1990,Ebeam_BackSc_2006},
on-wire lithography \cite{OnWireLitho_2005}, etc.  Interelectrode
distances down to 1-2~nm may be achieved
\cite{EM3_2002,ED2_2000,Ebeam_TEM_2007} by some of these methods, though
without much control of gap aspect ratio.

A significant challenge is fabricating two electrodes separated by a
nanometer gap running parallel over a macroscopic width
(high-aspect-ratio (HAR) nanogaps).  HAR nanogap fabrication has been
demonstrated based on a selective etching of cleaved GaAs/AlGaAs
heterostructures \cite{Tornow_2005,HARgap_2007}.  However, this method
requires particular substrates and allows only restricted gap
geometries.  A much simpler technique was proposed recently
\cite{Wind2_2006} with potentially no limitations on the width of the
gap.  Two separate lithographic patterning steps are used to define
first and second electrodes, while the interelectrode separation is
controlled by the oxidation of an Al sacrificial layer deposited upon
the first electrode. The native aluminum oxide layer, Al$_{2}$O$_{3}$,
overhangs the underlying metal and serves as a mask during the
deposition of the second electrode.  This layer must be removed
afterwards, but since Al$_{2}$O$_{3}$, corundum, is one of the most
chemically inert materials \cite{EtchReview_2003}, removal by direct
chemical etching is very difficult.  In a refinement, the authors
deposited an additional sacrificial layer of SiO$_{2}$ and
subsequently used etchant for SiO$_{2}$ to remove the SiO$_{2}$ and
Al/Al$_{2}$O$_{3}$ layers \cite{Wind2_2006}.  The use of SiO$_{2}$
etchant greatly limits the use of this approach in conjunction with
conventional silicon electronics.  While this method potentially
allows fabrication of HAR nanogaps, the reported aspect ratios are
less than 10 \cite{Wind2_2006}.

In this letter we report a highly reproducible and flexible method for
nanometer-sized (10-20~nm) gap fabrication with aspect ratios
exceeding 1000.  Modifying the original method \cite{Wind2_2006} by
replacing the Al layer with Cr results in a significantly more
flexible process.  A recipe is presented for effectively unlimited
aspect ratios with nearly 100\% yield.  Fabricated nanogap devices
were used for nanoscale electrical characterization of magnetite thin
films.  A characteristic conductance phenomenon related to magnetite
thin films\cite{Our_magnetite_2007} was observed indicating that
electronic transport is dominated by the nanometer region between
electrodes and there are no cross-gap shorts along gaps tens of $\mu$m
wide.  These nanogaps allow studies of this conductance switching at
voltages lower than those required in magnetite devices made with
standard lithography.

\begin{figure}[h!]
\begin{center}
\includegraphics[clip,width=7cm]{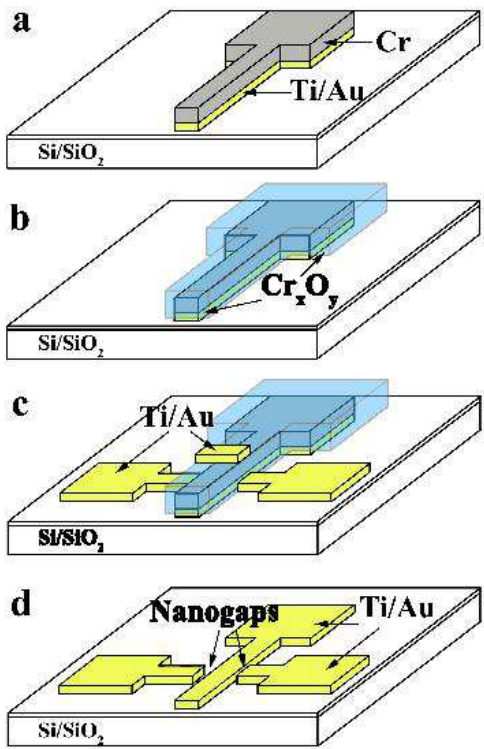}
\end{center}
\vspace{-5mm}
\caption{\small Schematic of the fabrication process (not to scale) (a) First electrode fabrication and deposition 1~nm Ti, 15~nm Au and 25~nm Cr layers. (b) Oxidation of Cr layer in ambient conditions giving an oxide layer a few nm thick (c) Second electrode patterning and deposition, with chromium oxide acting as a mask. (d) Etching away Cr/Cr$_{x}$O$_{y}$ and overlying second-step Ti and Au layers.}
\label{fig1}
\end{figure}

Electrode fabrication consists of three main steps
outlined in Fig.~\ref{fig1}.  Starting substrates were either silicon
wafers with 200~nm thermally grown SiO$_{2}$ layer, or 40-60~nm thick
magnetite films on MgO substrates \cite{Schvets_JAP_2004}.
Poly-(methylmethacrylate) (2\% in chlorobenzene) was spin-coated at
3000~rpm on the wafers to form a 500~nm resist layer, followed by
baking at 180~$^{\circ}$C for 2~min.  Conventional electron beam
lithography (JEOL 6500, 30kV) was used to define the first electrode.
After e-beam exposure the pattern was developed in a 1:3 volume
mixture of methyl-isobutyl-ketone (MIBK) and isopropanol (IPA) at room
temperature (RT).  Metallization via electron beam evaporation of 1~nm
Ti (adhesion layer), 15~nm Au and 10-45~nm Cr layers followed, with
subsequent lift-off in acetone to complete the first electrode
(Fig.~\ref{fig1}a).

Exposure to ambient conditions results in an oxide layer
(Cr$_{x}$O$_{y}$) (Fig.~\ref{fig1}b) that overhangs the edges of the
first Ti/Au electrode.  A second Ti/Au electrode is then fabricated by
e-beam lithography with only rough micrometer alignment of the second
electrode relative to the first one.  During the deposition of the
second electrode 1~nm Ti (0.1 nm/sec) and 15~nm Au (0.2 nm/sec)
layers, the Cr$_{x}$O$_{y}$ oxide on top of the first Ti/Au electrode
protects an few-nm-wide strip of substrate around the first Ti/Au
electrode (Fig.~\ref{fig1}c).  The suggested ``double pad'' geometry
of the second-step electrode (Fig.~\ref{fig1}c) simultaneously
produces two addressable gaps in one step.  Varying the source-sample
distance (30-45~cm) and deposition rates (0.1-0.2~nm/s) had no effect
on the results.

Finally the wafer was placed into chromium etchant solution (CR-7,
Cyantek) at room temperature for 1.5 min. to remove the
Cr/Cr$_{x}$O$_{y}$ layer and overlying second-step Ti and Au layers on
top of the first-step electrode.  As a result two pairs of electrodes
are produced, each pair separated by a nanogap defined by the
thickness of Cr$_{x}$O$_{y}$ layer (Fig.~\ref{fig1}d).  For fixed Cr
thickness, the aspect ratio of the gap is set by the width of the
second-step electrode and can be varied over a large range; devices
with 10-20~nm gaps 10-20~$\mu$m wide were successfully fabricated.
The overall yield of the process (unshorted devices) is limited by the
etching yield, and approaches nearly 100\% when Cr layer thickness is
optimized.    

\begin{figure}[h!]
\begin{center}
\includegraphics[clip, width=6cm]{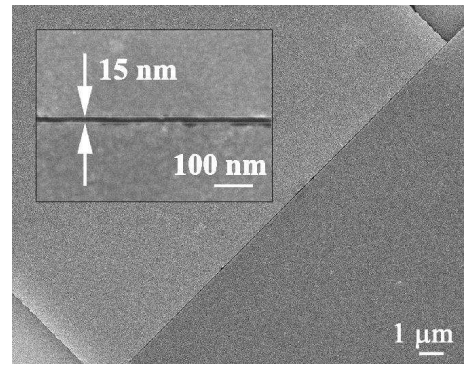}
\end{center}
\vspace{-5mm}
\caption{\small Scanning electron microscopy image of two gold electrodes on Si/SiO$_{2}$ wafer separated by 10-20~nm gap running along 20~$\mu$m (rotated by 45 deg.). The inset shows high magnification SEM image of a small fraction of the nanogap.}
\label{fig2}
\end{figure}

\begin{figure}[h!]
\begin{center}
\includegraphics[clip, width=8cm]{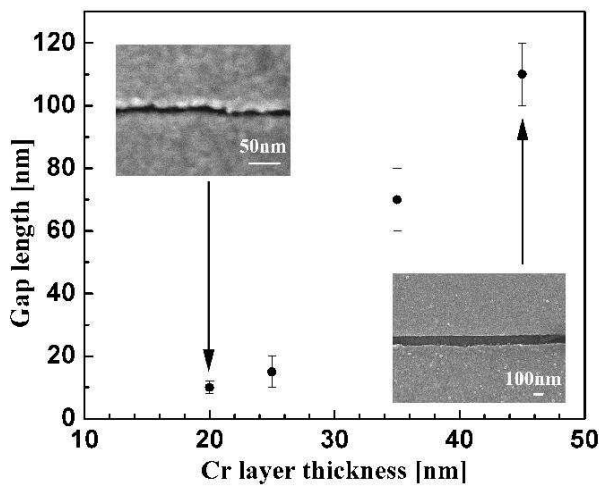}
\end{center}
\vspace{-5mm}
\caption{\small Dependence of the gap size on the thickness of deposited Cr layer. Each point represents an average over 16 devices fabricated under the same conditions. Insets show SEM images of produced gaps for 20~nm- (top) and 45~nm-thick (bottom) Cr layer.}
\label{fig3}
\end{figure}

\begin{figure}[h!]
\begin{center}
\includegraphics[clip, width=7.5cm]{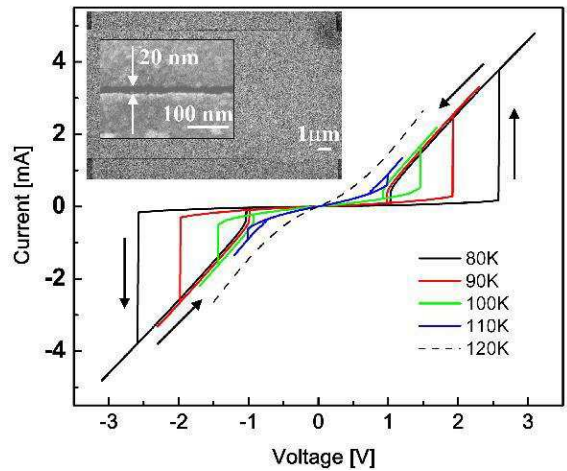}
\end{center}
\vspace{-5mm}
\caption{\small Current-voltage characteristics at different temperatures for a device produced on 50~nm magnetite film. Arrows indicate the direction of voltage sweeping. Inset shows SEM images of the electrodes and gaps on magnetite film.}
\label{fig4}
\end{figure}

Representative scanning electron micrographs of such electrodes on a
Si/SiO$_{2}$ wafer and Fe$_{3}$O$_{4}$ film on MgO substrate are shown
on Fig.~\ref{fig2} and Fig.~\ref{fig4} inset, respectively.  The gap
separation is 10-20 nm for electrodes on Si/SiO$_{2}$ wafer and
$\sim$20 nm for the magnetite film.  This separation (10-20~nm) is
much larger than the expected thickness of native amorphous
Cr$_{x}$O$_{y}$ formed at ambient conditions ($\sim$2~nm
\cite{Cr_native}).  Increasing the thickness of the Cr layer leads to
wider gaps (Fig.~\ref{fig3}), also inconsistent with the formation of only
native oxide.   

While the precise mechanism is still under investigation, two factors
likely contribute to this dependence on Cr thickness.  First, partial
thermal oxidation of the Cr layer may increase the oxide thickness,
since during the fabrication procedure the chromium layer undergoes
two heating processes: an elevated temperature in the evaporator
chamber during the evaporation process, and
PMMA baking for second step lithography
(180~$^{\circ}$C, 2 min).  Thermal oxide growth of
Cr$_{x}$O$_{y}$ primarily takes place through outward cation
diffusion; thus new oxide forms mainly on top of the existing
scale\cite{Cr_Mech}, increasing the resulting oxide layer overhang and
gap size.  Second, internal stresses generated within the (primarily)
Cr layer due to inward diffusion of oxygen\cite{Cr_stress} may lead to
elastic deformation of the underlying Cr.  Shear stresses at the
Cr/Cr$_{x}$O$_{y}$ interface would produce a deformation of the Cr
layer that would favor increased gap size and would increase with Cr
layer thickness, as observed in the experiment (Fig.~\ref{fig3}).
{\it Complete} oxidation of the Cr layer is unlikely, since
significant oxidation of Cr typically requires temperatures exceeding
300~$^\circ$C\cite{Graham_2002}.

Significantly decreasing the thickness of the Cr layer to achieve
narrow gaps increases the threat of accidental connections bridging
the gap.  An image of the gap produced by means of a 15-20~nm thick
sacrificial Cr layer shows that the gap size is well below 10~nm
(Fig.~\ref{fig3} inset), but the electrodes are not completely
separated from each other with some minor connections being clearly
observed in larger images. The gap edge of the second-deposition
electrodes tends to be more rough than that of the first-deposition
electrodes because of the morphology\cite{Cr_native}  of the Cr$_{x}$O$_{y}$ film.

The thickness of the Cr layer also appears to be critical in the
etching process; when the Cr layer thickness is below 10~nm, the wet
etching consistently resulted in the overlying second-deposition Ti
and Au layer fragments landing on the gap, shorting the two
electrodes.  Presumably, the etching rate of chromium layer is so
rapid that the overlying Ti/Au layer sticks to the first electrode
before it can be washed away.  Thus, the optimum thickness of the Cr
layer is 25~nm to assure the reliable formation of narrowest but at
the same unshorted gaps (Figs.~\ref{fig3},~\ref{fig2}).


Electrodes produced with this technique on Si/SiO$_{2}$ wafers were
electrically tested for shorting at room temperature (300K).  Upon
voltage sweeps from 0~V to 0.1~V more than 90\% of devices showed
currents below the measurable limit (pA), indicating an interelectrode
resistance exceeding tens of G$\Omega$.  This proves complete
isolation of the two electrodes without any connections or gap
irregularities narrow enough ($\sim$~1-2~nm) to allow tunneling.

We apply this fabrication process to examine the electronic properties
of thin (40-60~nm) single-crystal magnetite films grown on MgO
substrates \cite{Schvets_JAP_2004}.  Magnetite undergoes a first-order
phase transition when cooled through $\sim$122~K, known as the Verwey
temperature, $T_{\rm V}$, with a structural deformation accompanied by
a drastic decrease in electrical conductivity \cite{Verwey_1947}.
Above T$_{\rm V}$ magnetite behaves as a ``bad metal'' whereas below
T$_{\rm V}$ insulating behavior is observed.  In recent nanoscale
transport experiments, magnetite nanostructures (both nanocrystals and
single-crystal thin films with lithographically defined electrodes)
were shown to exhibit a sharp, voltage-driven transition in their
electrical conduction at temperatures below $T_{\rm V}$
\cite{Our_magnetite_2007}.  At a given temperature the transition
voltage is observed to decrease with decreasing interelectrode gap
size; thus HAR nanogaps are ideal structures for examining this
transition.

As a representative example, Fig.~\ref{fig4} shows current-voltage
characteristics of 50nm-thick film at different temperatures.  Sharp
conduction transitions hysteretic in voltage below 120~K are evident.
Upon approaching a temperature-dependent switch-ON voltage, the
conductance (current) jumps 1-2 orders of magnitude, nearly
approaching the conductance of the high-temperature ``metallic'' state
of magnetite.  The switching voltages increase as temperature
decreases while conductance after the transition remains essentially
the same.  The devices enabled by the HAR fabrication technique allow
us to achieve the necessary electric field strengths for switching at
relatively modest voltages (below 1V at 110K (Fig.~\ref{fig4}),
reducing the risk of sample damage and heating.  The HAR nanogaps lead
to uniform electric fields across the channel.

Fabrication of magnetite devices with the required small gap spacings
via the Cr HAR method has been much more reliable and consistent than
attempts to do so using two-step e-beam lithography.  Furthermore, the
survival of the relatively sensitive Fe$_{3}$O$_{4}$ film during this
process shows the chemical selectivity and versatility of the Cr-based
technique. In addition to Ti/Au metallization our technique was
successfully applied to fabricate Pt, Al, Fe and Ti electrodes on
Si/SiO$_{2}$ wafers.  The HAR closely-spaced electrodes enabled by
this approach should find broad applicability in nanoscale electronic
measurements.

This work was supported by the US Department of Energy grant
DE-FG02-06ER46337.  DN also acknowledges the David and Lucille Packard
Foundation and the Research Corporation, and Z.K. Keane and C. Slavonic
for valuable contributions in the early stages of this work. RGSS and
IVS acknowledge the Science Foundation of Ireland grant 06/IN.1/I91.




\end{document}